\documentclass[psfig]{mn2e}
\def\ltsima{$\; \buildrel < \over \sim \;$}
\def\simlt{\lower.5ex\hbox{\ltsima}}
\def\gtsima{$\; \buildrel > \over \sim \;$}
\def\simgt{\lower.5ex\hbox{\gtsima}}
\input{epsf}

\title{Gravitational instability in binary protoplanetary disks; new constraints
on giant planet formation}

  \author{Lucio Mayer $^1$, James Wadsley$^2$, Thomas Quinn$^3$, Joachim 
Stadel$^1$ \\$^1$Institute of Theoretical Physics, 
University of Z\"urich, Winterthurerstrasse 190, 8057 Zurich, Switzerland
\\$^2$Department of Physics \& Astronomy, McMaster University, 1280 Main St. 
West, Hamilton ON L8S 4M1 Canada\\$^3$Department of Astronomy, University of 
Washington, Seattle, WA 98195, USA}

\begin{document}

\maketitle

\begin{abstract}

We use high resolution 3D SPH simulations to study the evolution of 
self-gravitating binary protoplanetary disks. Heating by shocks
and cooling are included. We consider different orbital separations and 
masses of the disks and central stars. Isolated massive disks ($M \sim 0.1 M_{\odot}$) 
fragment into protoplanets as a result of gravitational instability for
cooling times comparable to the orbital time. Fragmentation does not occur  
in binary systems with a separation of about 60 AU.
This is because efficient heating owing to strong tidally induced spiral shocks
damps any overdensity.
The resulting temperatures, above 200 K, would vaporize water ice in the outer
disk, posing a problem even for the other model of giant planet formation,
core-accretion. Light disks ($M \sim 0.01 M_{\odot}$) do not fragment  
but remain cold because their low self-gravity inhibits strong shocks.
Core accretion would not be hampered in the latter.
At separations of about 120 AU the efficiency of
fragmentation by disk instability rises and approaches that in  
isolated  systems. If disk 
instability is the main  formation mechanism for giant planets, 
on going surveys targeting binary systems should find 
considerably fewer planets in systems with separations 
below $100$ AU.

\end{abstract}


\section{Introduction}

The recent discovery of extrasolar planets (Mayor \& Queloz 1995) has 
ignited renewed interest
in models of giant planet formation. In the conventional model, core accretion 
(Lissauer 1993), it is difficult to grow planets of several Jupiter masses
in less than a few million years, the typical disk lifetime estimated from
observations (Haisch, Lada \& Lada 2001). This problem is exacerbated by
the fast inward migration rates produced by the disk-planet interaction as
well as by the low accretion rates ensuing once a planet is big enough to
open a gap (Nelson et al. 2001; Bate et al. 2003; Nelson \& Benz 2003). 
Consequently, the disk instability model, in which giant planets
arise in only a few disk orbital times (less than a thousand years) 
from the fragmentation of a massive, gravitationally unstable disk
(Boss 1997; 2002; Pickett 2000;2003) has gained new attention (Mayer et al. 2002; Rice et al. 2003a,b).

The majority of solar-type stars in the Galaxy belong to double or multiple
stellar systems (Duquennoy \& Mayor 1991; Eggenberger et al. 2004).
Binaries can be 
formed by the fragmentation of a single bar-unstable
molecular cloud core into two distinct objects (Boss 1986; Burkert,
Bate \& Bodenheimer 1997), from the collision of two dense cores in a
giant molecular cloud (Whitworth et al. 1995)  
or owing to the capture of neighboring stars in dense star 
forming regions (Bate et al. 2002). Fragmentation  
is usually considered as the main channel of binary formation and can take 
place in any type of environment (Horton, Bate \& Bonnell 2001).
In such scenario two star-disk systems should form 
if the initial separation is larger than $10$ AU, while at smaller
separations a circumbinary disk can arise (Bate 2000).
Radial velocity surveys have
shown that planets exist in some binary or multiple stellar systems
where the stars have separations from 20 to several thousand AU 
(Eggenberger et al. 2004). Although the samples are still
small (20 out of the 120 known extrasolar planets are in binary
systems), attempts have been made to compare properties
of planets in single and multiple stellar systems 
(Patience et al. 2003; Udry et al. 2004). The first adaptive optics surveys
designed to quantify the relative frequency of planets
in single and multiple systems are just starting (Udry et al. 2004). 
These surveys could offer a new way to test theories of giant planet formation 
provided that different models yield different predictions as for the
effect of a stellar companion.

So far, two works have studied giant planet formation in binary systems.
Both focused on the disk instability model and reached opposite 
conclusions. Nelson (2000) performed 2D SPH simulations of protoplanetary
disks that did not form protoplanets in isolation due to quite long 
cooling times (Johnson \& Gammie 2003) and found fragmentation to be even 
more unlikely in the presence of another disk with identical mass at a
mean separation less than 100 AU. In fact the disks were achieving a high
stability owing to heating by shocks occurring along the spiral arms.
Boss (1998, 2003) performed 3D grid calculations of a disk interacting with
a stellar companion also closer than 100 AU and found that giant planet 
formation by disk instability can be enhanced by the tidal perturbation.

In this Letter we revisit giant plant formation in a binary 
system using high resolution 3D SPH simulations. 
In Mayer et al. (2002; 2004),
we have shown that a high resolution is crucial in order to
decide whether a massive protoplanetary disk will fragment into long-lasting
clumps as a result of gravitational instability.
Fragmentation requires the disk to cool on a timescale
comparable to or shorter than the disk orbital time 
(Rice et al. 2003a,b; Mayer et al. 2003). 
Here we consider binary systems of protoplanetary disks which
mostly fragment in isolation (Mayer et al. 2003) and investigate
whether the tides raised by the companion enhance or suppress
fragmentation.

\section{Models and Simulations}

The initial conditions comprise two protoplanetary disks, usually
of equal mass, orbiting around each other. Each disk orbits a central 
star as well. 
In a binary system arising from the fragmentation of a molecular
cloud core one expects the two disks to form in the same plane and 
be rotating in the same direction relative to their orbital motion 
(e.g. Bate 2000).In this paper we will
restrict our investigation to such coplanar prograde configurations 
(see also Nelson 2000).
Table 1 lists the most important parameters of the simulations.
The disks extend from 4 to 20 AU and their orbital separation is
in the range 60-120 AU (semi-major axis $a=29$ or $58$ AU, eccentricity $e
= 0.14$), corresponding to some of the smallest separations among
binary systems with detected giant planets in radial velocity surveys
(Eggenberger et al. 2004).
Each disk is represented by 200000 SPH particles with
equal mass and fixed gravitational softening of $0.06$ AU, while the central 
star has a softening of 2 AU (see Mayer et al., 2004, hereafter MA04).
The initial temperature reaches about $400$ K at the inner boundary and levels 
off at $65$ K close to 10 AU. The details
of the temperature and surface density profiles of the disks can be found
in MA04. 

\begin{figure}
\epsfxsize=9truecm
\epsfbox{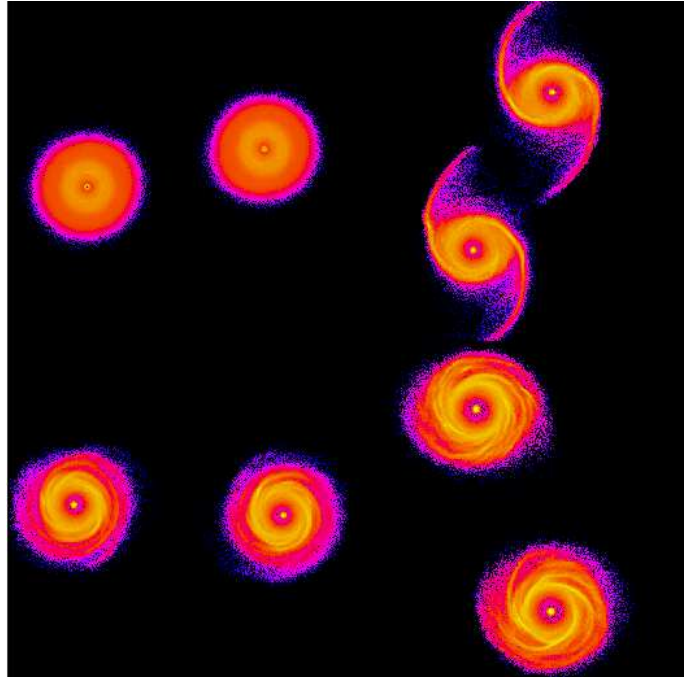}
\caption{Color coded projected gas density in the plane of
the binary orbit (brighter colors correspond to higher densities)
for run RB1a. Densities between $10^{-15}$ and $10^{-8}$ g/cm$^3$ are shown.
Boxes are about 100 AU on a side. From top-left to bottom-right, 
snapshots at 16 years (shortly after the beginning),
150 years (after first pericenter passage), 300 years and 450 years (close
to the second pericenter passage) are shown. See  text for details.
Note the overdensities along the spiral arms at 150 years; they
are rapidly quenched by the high pressure along the arms.}
\end{figure}

Disk models are in equilibrium when evolved for several orbital
times in absence of self-gravity (see MA04 for details). Models 
as light as the minimum mass solar nebula ($0.012 M_{\odot}$) and as massive
as the heaviest among T Tauri disks (D'Alessio et al. 2001) are considered.
The shape of the profile of the Toomre Q parameter is the same in all
disks, while its normalization varies  according to the mass of the disk.
We recall that a gaseous (nearly) keplerian disk is stable
against local axisymmetric perturbations if $Q > 1$, $Q=\Omega c_s/
\pi G \Sigma$, where $\Omega$ is the angular frequency, $c_s$ is 
the sound speed, $G$ is the gravitational constant and $\Sigma$ is 
the surface density of the disk.
The initial minimum Q parameter (reached near the disk edge) is 
$Q_{min} \sim 1.4$ or higher (see Table 1), hence close or 
above the threshold for fragmentation 
in isolation (MA04; Johnson \& Gammie 2003). 

The radiative cooling time is taken to be proportional to the 
local orbital time, following Rice et al. (2003a). 
Inside 5 AU cooling is switched off in order
to maintain temperatures high enough to be comparable to those in 
protosolar nebula models (e.g. Boss 1998).
Cooling is also switched off in regions reaching 
a density above $10^{-10}$ g/cm$^3$ to account 
for the local high opacity; the choice of the density threshold 
is motivated by the simulations of Boss (2002) with radiative transfer
in which the temperature of the gas is observed to  evolve nearly 
adiabatically above such densities. 
We consider cooling times going from $0.3$ to $1.5$ the local orbital time. 
The jury is still out on whether the cooling times adopted here 
are credible or excessively short, but recent calculations by
Boss (2002) and Johnson \& Gammie (2003), which use different approximate 
treatments of radiative transfer, do find cooling times of this magnitude
through a combination of radiative losses and convection (but see Mejia
et al. 2003, 2004).
In any case, here we are 
interested in comparing the outcomes of isolated and binary systems for 
the same choice of the cooling time, in particular when cooling is 
strong enough to lead to fragmentation in isolated disks.
Heating by compressions and shocks is included in the simulations.
We usually adopt $\gamma=7/5$, appropriate for molecular hydrogen, 
with a few simulations having  $\gamma = 5/3$. 
Shocks are modeled using the standard Monaghan viscosity with $\alpha=1$ and 
$\beta = 2$ plus the Balsara correction term that removes unwanted shear
viscosity (see Wadsley, Stadel \& Quinn 2003).
The full analysis of the runs with isolated disks is carried out elsewhere 
(Mayer et al. 2003; Mayer et al., in preparation).
Here we only report on whether the isolated disks formed 
clumps or not (see Table 1).

\begin{table*}
\centering
\caption{Parameters of the simulations.
Column 1: Name of run.
Column 2: Disk mass (A)($M_{\odot}$). 
Column 3: Disk mass (B)($M_{\odot}$).
Column 4: Star mass (A)($M_{\odot}$.)
Column 5: Star mass (B)($M_{\odot}$.)
Column 6: Semi-major axis of the orbit (AU).
Column 7 : Initial minimum Toomre Q parameter.
Column 8: Cooling time (in units of the orbital time).
Column 9: $\gamma$.
Column 10: Whether the disks in a binary fragments or not ("tr" stands 
for transient  clumps).
Column 11: Whether the isolated disks fragment or not ("tr" stands for transient clumps).
}
\begin{tabular}{lcccccccccc}
Model &  $M_{dA}$ & $M_{dB}$ & $M_{*A}$& $M_{*,B}$ & $a$ &  $Q_{min}$ & 
$t_{cool}$ & $\gamma$ & clumps (bin) & clumps (is) \\
%\hline 

RB1a          & 0.1 & 0.1 & 1 & 1 & 29 & 1.4 & 0.3 & 1.4 & no & yes  \\
RB1b          & 0.1 & 0.1 & 1 & 1 & 29 & 1.4 & 0.5 & 1,4 & no & yes \\
RB1c          & 0.1 & 0.1 & 1 & 1 & 29 & 1.4 & 1 & 1.4 & no & yes \\
RB1d          & 0.1 & 0.1 & 1 & 1 & 29 & 1.4 & 1.5 & 1.4 & no & yes  \\
RB1e          & 0.1 & 0.1 & 1 & 1 & 29 & 1.4 & 0.3 & 1.66 & no & yes \\
RB1f          & 0.1 & 0.1 & 1 & 1 & 29 & 1.4 & 1 & 1.66 & no & yes \\
RB2a          & 0.05 & 0.05 & 1 & 1 & 29 & 2.8 & 0.5 & 1.4 & tr & no \\
RB2b          & 0.05 & 0.05 & 1 & 1 & 29 & 2.8 & 0.3 & 1.4 & yes & no \\
RB3a          & 0.08 & 0.08 & 1 & 1 & 29 & 1.75 & 0.5 & 1.4 & no & yes \\
RB3b          & 0.08 & 0.08 & 1 & 1 & 29 & 1.75 & 0.3 & 1.4 & yes & yes \\
RB4a          & 0.012 & 0.012 & 1 & 1 & 29 & 11 & 0.3 & 1.4 & no & no\\
RB4b          & 0.012 & 0.012 & 1 & 1 & 29 & 11 & 1.5 & 1.4 & no & no\\
RBm2          & 0.1 & 0.05 & 1 & 0.5 & 29 & 1.4 (2) & 0.3 & 1.4 & tr & yes \\
RBwa          & 0.1 & 0.1 & 1 & 1 & 58 & 1.4 & 0.3 & 1.4 & yes & yes \\
RBwb          & 0.1 & 0.1 & 1 & 1 & 58 & 1.4 & 0.5 & 1.4 & yes & yes \\
RBwc          & 0.1 & 0.1 & 1 & 1 & 58 & 1.4 & 1 & 1.4 & tr & yes \\
RBwd          & 0.1 & 0.1 & 1 & 1 & 58 & 1.4 & 0.5 & 1.66 & tr & yes \\

\end{tabular}
\label{t:simul}
\end{table*}

\section{Results}

We begin by describing the outcome of the runs with the smallest 
orbital separations, corresponding to $a=29$ AU (see Table 1). 
Figure 1 shows the evolution of the  disks in one of such simulations.
Disk models are typically followed for two orbits. A binary orbit corresponds
to  about 288 years, or, 
equivalently, 10 disk orbital times at 10 AU from the disk center.
Two calculations were extended further for two more orbits, but, since
gravitational instability acts on the disk orbital timescale,
fragmentation should occur during the first binary orbit.
Disks start at apocenter and develop a
strong two-armed spiral pattern after crossing pericenter. The strong
non-axisymmetric torques redistribute mass and angular momentum in the 
disks, which soon 
develops higher order spiral modes  at distances between 12 and 15 AU
from their centers. 
In models that do not fragment (see Table 1) transient, 
moderately strong high order spiral arms continue to develop at 
subsequent pericentric passages, but $Q_{min}$ always remains above 
the threshold  for stability (see Figure 2).
In models that fragment, $Q_{min}$ drops below 1 (see Figure 2), and 
clumps appear on the disk side which is 
further from the other disk, along a strong unwinding trailing spiral arm. 
On the other side of the disk the developing overdensities are 
destroyed as tides tear them apart.

We can identify three regimes as for the disk response
to the tidal perturbation. Such regimes can be distinguished 
based on the disk mass. Very light disks,
which are completely stable in isolation, develop a clear two-armed
spiral pattern but their self-gravity is too low to amplify the waves
and sustain the instability (Figure 3).
This spiral mode 
simply evolves periodically with the orbit, strengthening at pericenter 
and weakening at apocenter.
Disks at the high mass end, which are strongly self-gravitating, and, when
isolated, undergo fragmentation into protoplanets for cooling times 
of order the orbital time (Mayer et al. 2003), develop a very strong 
and complex spiral pattern after the first pericenter passage, with 
a dominant two-armed mode (see Figure 1).
However, such strong amplification of the
spiral waves also leads to intense compressional heating along the arms.
These spiral shocks increase
the temperature of the outer disk by nearly a factor of 3 and bring the 
system towards stability (Figure 2). 
This is why the spiral arms fade away considerably during the 
second orbit (see Figure 1).
The suppression of fragmentation by tidally induced shock heating
was also advocated by Nelson (2000).
In the intermediate mass regime, $M_d =0.05,0.08 M_{\odot}$, self-gravity is 
non-negligible and spiral instabilities are visibly amplified, but shock 
heating is mild enough for non-axisymmetric features to last longer and build
up more pronounced overdensities (Figure 3). These disks can fragment for 
the shortest among the cooling times considered here despite the fact
that they can avoid fragmentation in isolation. Such short cooling times are 
probably unrealistic, nevertheless it is interesting that there are
cases in which tides can increase the susceptibility to 
fragmentation as found by Boss (1998).

\begin{figure}
\begin{center}
\epsfxsize=6truecm
\epsfbox{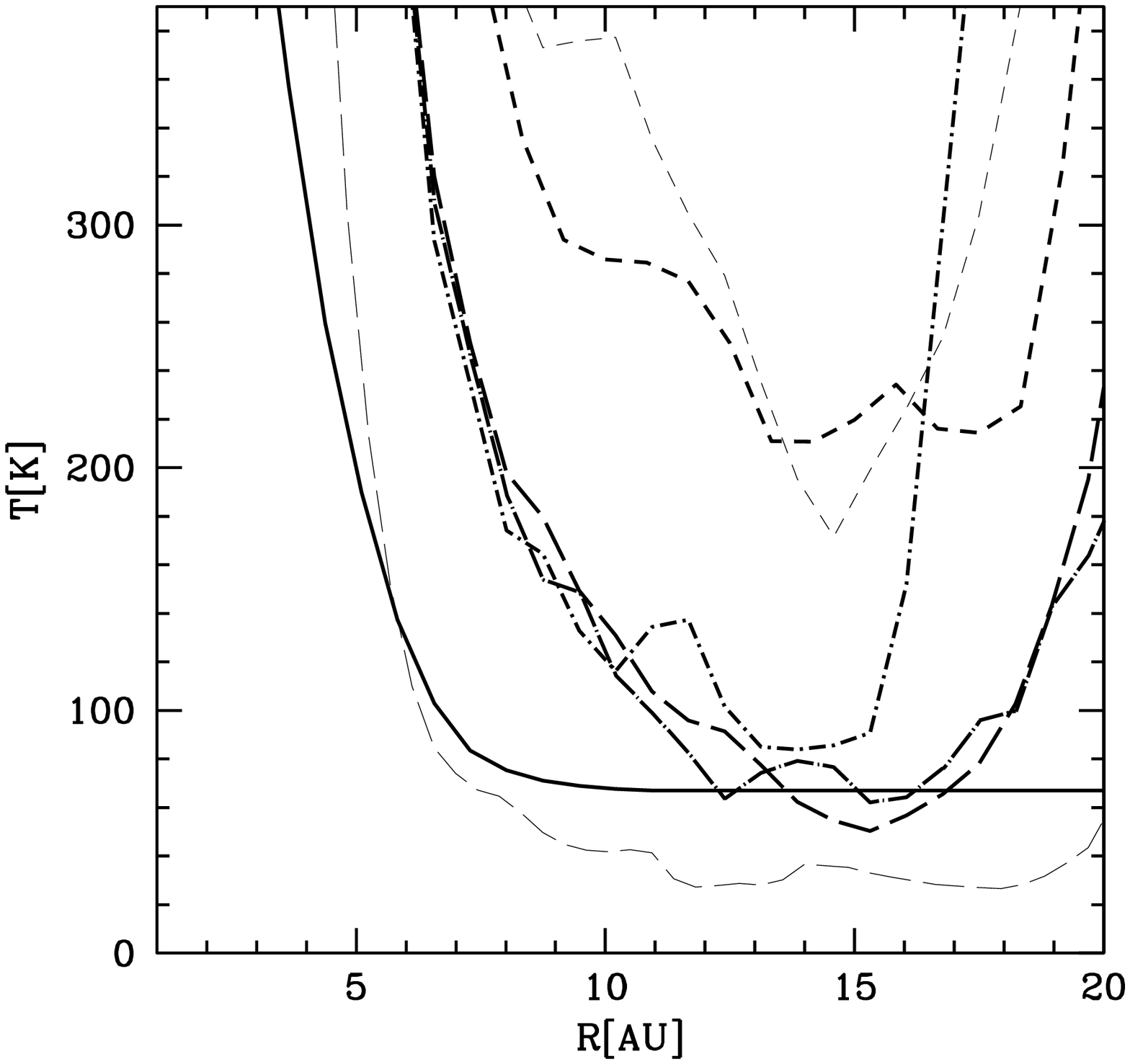}
\epsfxsize=6truecm
\epsfbox{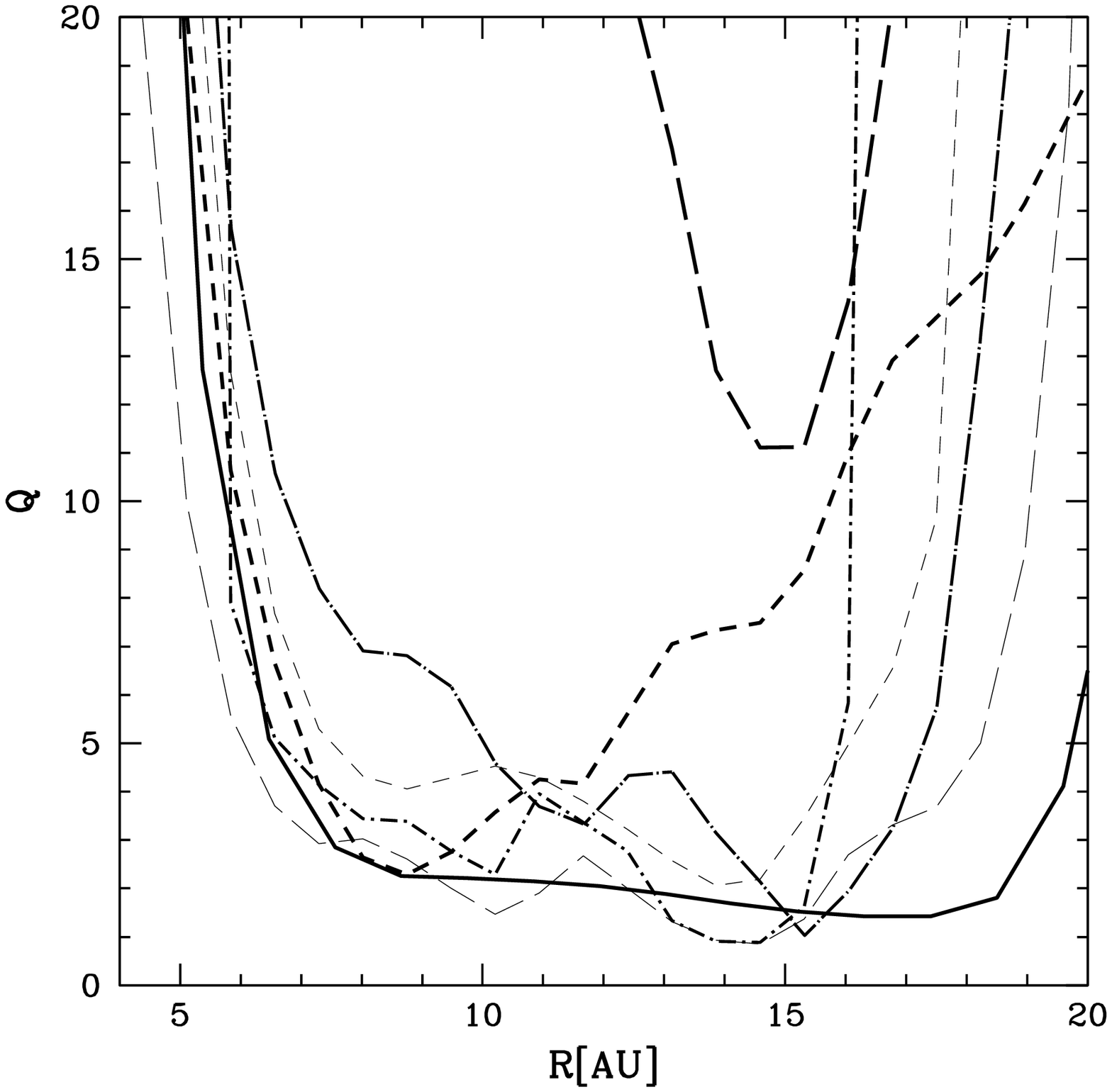}
\caption{Azimuthally averaged midplane temperature (top) and 
$Q$ (bottom) profiles  
at the time of maximum amplitude of the overdensities
(at between 120 and 200 years depending on the model). 
Disks with outer temperatures above 100 K are too hot
to fragment, as shown by the high values of $Q$.
We show: initial conditions  (thick solid line, Q profile normalized 
as in the RB1 runs),
RB1a (thick short-dashed line), RB1d (thin short-dashed line), isolated 
disk run with model used in RB1a (thin long dashed line), RBwb (thick 
dotted short dashed line), RB4b(thick long dashed line), RB2b (thick dotted
long dashed line).}
\end{center}
\end{figure}

We can ask how realistic are the temperatures seen in our
simulations. The high temperatures developing in the outer parts of the 
most massive disk models are comparable to those in the simulations 
of Nelson (2000).
This author calculated the corresponding radiation flux at far-infrared and
radio wavelengths (from 870 $\mu$m to 1.3 cm)
assuming that the disk emits like a blackbody and obtained fluxes somewhat
lower than observed in a prototype young binary protostellar disk system,
L1551 IRS 5 (Bachiller, Tafalla \& Cernicharo 1994).
Therefore the temperatures seen in our simulations are
probably a conservative estimate of those occurring in
real binary systems.  

Temperature in excesse of 200 K as those obtained here (Figure 3) for the most 
massive
disk models (especially along the spiral shocks)
would be enough to vaporize water ice. The latter should contribute 
almost half of the mass
of solid material in a protoplanetary disk (Pollack et al. 1994), and
is therefore a fundamental building block of large solid grains and, 
ultimately,
of planetesimals. A reduced growth of rocky planetary embryos could result,
and therefore giant planet formation by core accretion could also
be less likely in such binary systems relative to isolated system (see also Nelson 2000).
Light or intermediate mass disks, instead, maintain outer disk temperatures
lower than 100 K between 10 and 20 AU (Figure 2) posing no problem for 
core accretion.

Many binary systems comprise stars with unequal masses. We 
simulated the interaction between two disk+star systems with masses
differing by a factor of 2 (run Rbm2, see Figure 3).
The most massive of the two disks ($0.1 M_{\odot}$), which never 
fragmented when interacting with an equally massive disk (e.g. run RB1a,
RB1b, RB1c), now manages to produce two clumps of roughly one Jupiter mass. 
Yet these clumps are quickly dissolved as the pressure still overcomes 
self-gravity.
The lighter disk forms strong multi-armed spirals, probably resulting from
strong swing amplification (as expected given the lower mass of the central
star, see MA04). Although the temperature remains quite
low ($Q$ drops close to unity locally), nascent overdensities are 
apparently sheared away before they are able to fragment because of the strong
tidal field of the more massive companion.
We note that a $0.1 M_{\odot}$ disk does not produce permanent clumps 
both while interacting with an identical disk (e.g. run RB1a) and when
orbiting a companion system half of its mass as in run RBm2.
Coversely, in run RB3b a disk only 20\% lighter (see Table 1) does
give rise to long-lived clumps while interacting with a system of 
identical mass (all these runs employ the same cooling times).
This suggests that tidally induced spiral shocks become 
too strong above some critical value of the disk
self-gravity, and heating becomes dominant, despite the fact that the
strength of the external perturbation changes significantly.

Among the few known binary
systems with planets the majority have stars with a projected 
separation  above 100 AU.  
In runs RBw(a,b,c) we evolved the massive
disk models on orbits with a larger semi-major axis, $a = 58$ AU. In all 
these runs
we wintnessed fragmentation (Figure 3), although in run RBwc, in which 
the cooling
time was the longest, clumps did not survive for more than $\sim 3$ disk 
orbital times. We conclude that at these larger orbital separations disk 
instability proceeds similarly to the case of isolated disks
because tidal forces are considerably weaker. Moreover, the temperatures
at such separations are low enough to guarantee the survival of ice grains
(Figure 1).

\begin{figure}
\begin{center}
\epsfxsize=8truecm
\epsfbox{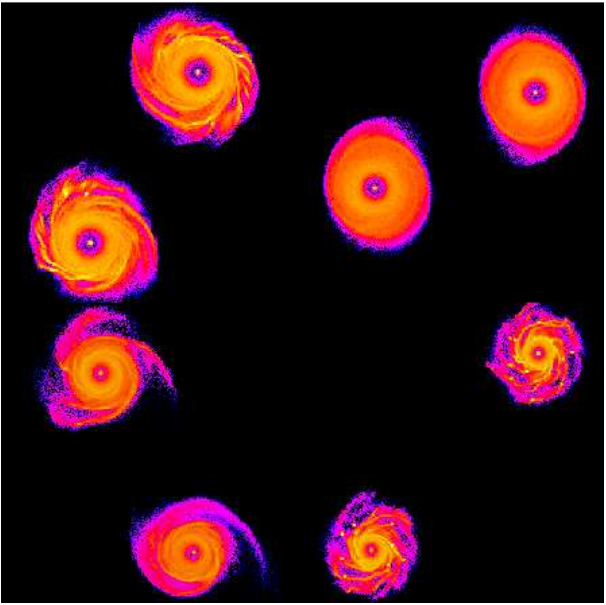}
\caption{Color coded projected gas density in the plane of
the binary orbit (see Figure 1).
We show selected snapshots from runs employing disks with different
masses.
From top-left to bottom-right, run RB2b, run RB4b, run RBm2 and run RBwb 
are shown at,  respectively, 200, 200, 140 years and 160 years 
(we have chosen time frames corresponding to the maximum growth of the 
overdense regions). Boxes are 100 AU (top), 130 AU (bottom left) and
200 AU (bottom right). The clumps seen in run RBm2 are
transient, while many of those in the runs RB2b and RBwb
survive and contract to densities $10^4$ times higher.}
\end{center}
\end{figure}

\section{Summary and Discussion}

We have shown that fragmentation by disk instability is suppressed 
in binary systems with orbital separations around 60 AU. This happens 
because shock heating overwhelms cooling and damps any
overdensity. Shock heating is stronger than in isolated disks because
the tidally forced spiral arms reach a much greater amplitude 
relative to spiral arms in isolated disks. This is especially true in 
the most massive disks since their higher self-gravity amplifies the 
spiral arms more efficiently.
The role of shock heating in disk instability has been
already recognized as crucial for understanding
if giant planets can form by disk instability (Pickett et al. 2000, 2003). 
In systems with a separation of 120 AU, disk temperatures remain quite low and
fragmentation proceeds more similarly to the isolated disks.
The high temperatures ($> 200$ K) developing in massive binary disk systems
with separations less than $60$ AU 
make it hard to form giant planets even by disk core accretion.
Intermediate mass systems are those in which both mechanisms are 
possible if cooling is very efficient whereas in binary disks with small 
masses, comparable to that of the minimum mass solar nebula model,
core-accretion is the only viable mechanism. Models of the
core accretion mechanism used to require a disk 3-4 times more massive 
than the minimum solar nebula in order to form Jupiter in less than 10 million
years (Lissauer 1993).
However, more recent models that account for orbital 
migration of rocky cores find formation timescales of a few million years
even in a minimum mass solar nebula since the cores feed more
efficiently with planetesimals as they migrate in the disk 
(Rice \& Armitage 2003; Alibert, Mordasini \& Benz 2004).
If core accretion can take place in light disks then giant planets could
form regardless of the presence or distance of a companion.
This suggests that binarity can be used to probe planet 
formation models. If the new surveys aimed at quantifying the
relative number of giant planets in single and binary systems 
will find no trend with binary separation disk instability cannot be 
the main formation mechanism. The opposite might be true if such trend emerges.

Our analysis was restricted to just one type of orbital configuration. 
Because of the low orbital eccentricity, the external perturbation
is nearly continous in amplitude.
This could favour high temperatures
through nearly continous compressional heating. Although Nelson (2000)
did not find any remarkable difference in systems with orbital 
eccentricities varying by a factor of 3, impulsive tidal perturbations, 
caused for example by a close fly by of a star or brown dwarf, which 
would be common in highly dynamical star formation scenarios 
(e.g. Bate et al. 2002), could produce a different outcome.
A strong shock would occur in this case but over time compressional
heating would be much lower.
We will invesigate such situations in a forthcoming paper. We will also
consider a larger variety of initial orbital configurations, for example
non-coplanar disks possibly resulting from a
capture event. 
The study of the geometry and relative orbits of debris disks around young binaries
will be needed to find out to what extent the simple orbital configurations used in
this paper are really representative.

\bigskip

We thank Stephane Udry, Anne Eggenberger and Michel Mayor for stimulating
discussions. L.M. thanks the
Swiss National Science Foundation for the financial support.  T.Q. is
supported by the NASA Astrobiology Institute. The simulations 
were performed on the Zbox supercomputer at the University of Zurich and on 
LeMieux at the Pittsburgh Supercomputing Center.

\end{document}